\documentclass[useAMS,usenatbib,usegraphicx]{mn2e}

\def\kelvin{\mbox{$\;$K}}
\def\perc2{\mbox{$\;$cm$^{-2}$}}
\def\percm2{\mbox{$\;$cm$^{-2}$}}

\def\ks{\mbox{$\;$ks}}              %kilosec: ksec%
\def\kms{\mbox{$\;$km s$^{-1}$}}       %km s -1%
\def\cms2{\mbox{$\;$cm s$^{-2}$}}      %cm s -2%
\def\ergss{\mbox{$\;$erg s$^{-1}$}}   %ergs s -1%

\def\arcsec{\rlap{$^{\prime\prime}$}\hbox to 2pt{}}
\def\arcmin{\rlap{$^{\prime}$}\hbox to 2pt{}}
\def\H0{\mbox{$\;$H$_{0}$}}
\def\hubble{\mbox{$\;$km$\;$s$^{-1}$Mpc$^{-1}$}}

\def\flux{\mbox{$\;$ergs$\;$cm$^{-2}\;$s$^{-1}$}}
\def\monolum{\mbox{$\;$ergs$\;$s$^{-1}\;$Hz$^{-1}$}}
\def\kev{\mbox{$\;$keV}}

\def\ev{\mbox{$\;$eV}}

\def\etal{et al.}
\def\gtorder{\mathrel{\raise.3ex\hbox{$>$}\mkern-14mu
             \lower0.8ex\hbox{$\sim$}}}
\def\ltorder{\mathrel{\raise.3ex\hbox{$<$}\mkern-14mu
             \lower0.8ex\hbox{$\sim$}}}
\def\apj{ApJ}
\def\apjl{ApJ}
\def\apjs{ApJS}
\def\mnras{MNRAS}
\def\aj{AJ}
\def\aap{A\&A}
\def\araa{ARA\&A}
\def\gca{Geochim.\ Cosmochim.\ Acta}

\begin{document}
\hyphenation{brems-strah-lung}

\title[X-ray Spectrum of FIRST J1556$+$3517]{The X-ray Spectrum and Spectral
  Energy Distribution of \\ FIRST J155633.8$+$351758: a LoBAL Quasar with a
  Probable Polar Outflow}

\author[R. C. Berrington \etal]{Robert
  C. Berrington,$^{1}$\thanks{rberring@bsu.edu} 
Michael S. Brotherton,$^{2}$
Sarah C. Gallagher,$^{3}$ 
\newauthor
Rajib Ganguly,$^{4}$ 
Zhaohui Shang,$^{5}$ 
Michael  DiPompeo,$^{2}$ 
Ritaban Chatterjee,$^{6,2}$ 
\newauthor
Mark Lacy,$^{7}$ 
Michael D. Gregg,$^{8,9}$ 
Patrick B. Hall,$^{10}$ \& 
S. A. Laurent-Muehleisen$^{11}$\\
$^{1}$Department of Physics and Astronomy, Ball State University, 
  Muncie, IN 47306, USA\\
$^{2}$Department of Physics and
  Astronomy, University of Wyoming, Laramie, WY 82071, USA\\
$^{3}$Department of Physics and Astronomy, University of Western
  Ontario, London, Ontario N6A 3K7, Canada \\
$^{4}$Department of Computer Science, Engineering, and Physics, University
of Michigan, Flint, Flint, MI  48502, USA\\
$^{5}$Tianjin Normal University, Tianjin 300387, China \\
$^{6}$Department of Physics, Presidency University, 86/1 College Street,
Kolkata 700073, WB, India
$^{7}$North American ALMA Science Center, National Radio Astronomy
  Observatory, Charlottesville, VA 22903-2475, USA\\
$^{8}$Physics Department, University of California, Davis, CA 95616, USA\\
$^{9}$Institute of Geophysics and Planetary Physics, Lawrence
Livermore National Laboratory, PO Box 808, L-413, Livermore, CA 94551-9900, USA\\
$^{10}$Department of Physics and Astronomy, York University,
Toronto, Ontario, M3J~1P3, Canada\\
$^{11}$Physics Division of Biological, Chemical, and Physical
Sciences, Illinois Institute of Technology, Chicago, Illinois 60616, USA}
\maketitle
%\email{rberring@uwyo.edu}
%\email{mbrother@uwyo.edu}
%\email{sgall@astro.ucla.edu}
%\email{ganguly@uwyo.edu}
%\email{shang@uwyo.edu}
%\email{mlacy@ipac.caltech.edu}

\begin{abstract}
We report the results of a new $60\ks$ {\it Chandra X-ray Observatory}
Advanced CCD Imaging Spectrometer S-array (ACIS-S)
observation of the reddened, radio-selected, highly polarized `FeLoBAL'
quasar FIRST J1556$+$3517. We investigated a number of models of varied
sophistication to fit the 531-photon spectrum.  These models ranged from
simple power laws to power laws absorbed by hydrogen gas in differing
ionization states and degrees of partial covering.  Preferred fits indicate
that the intrinsic X-ray flux is consistent with that expected for quasars of
similarly high luminosity, i.e., an intrinsic, dereddened and unabsorbed
optical to X-ray spectral index of $-1.7$.  We cannot tightly constrain the
intrinsic X-ray power-law slope, but find indications that it is flat (photon
index $\Gamma = 1.7$ or flatter at a $>99$\% confidence for a neutral hydrogen
absorber model). Absorption is present, with a column density a few times
$10^{23}\percm2$, with both partially ionized models and partially covering
neutral hydrogen models providing good fits.  We present several lines of
argument that suggest the fraction of X-ray emissions associated with the
radio jet is not large.

We combine our {\it Chandra} data with observations from the literature to construct
the spectral energy distribution of FIRST J1556$+$3517 from radio to X-ray
energies.  We make corrections for Doppler beaming for the pole-on radio jet,
optical dust reddening, and X-ray absorption, in order to recover a probable
intrinsic spectrum. The quasar FIRST J1556$+$3517 seems to be an intrinsically
normal radio-quiet quasar with a reddened optical/UV spectrum, a
Doppler-boosted but intrinsically weak radio jet, and an X-ray absorber not
dissimilar from that of other broad absorption line quasars.  
%Interestingly,
%it does seem to be deficient in the far-infrared part of the spectrum.
\end{abstract}


\section{INTORDUCTION}
\label{sec:introduction}

A substantial fraction of quasars possess intrinsic high-velocity outflows
along the line of sight, the most extreme of which are characterized by broad
absorption lines (BALs): broad, blueshifted resonance absorption lines seen in
the rest-frame ultraviolet.  The dynamics of these intrinsic outflows appear
to be the result of radiative acceleration (e.g., Arav, Korista, \& Begelman
1995; Ganguly \etal\ 2007; DiPompeo \etal\ 2012b).  Taken at face value, the
ultraviolet BALs suggest absorbing column densities of $N_{H} \sim
10^{20}$--$10^{21}\perc2$ in these outflows \citep{hamann:93}, although there
is evidence that the actual column densities are much higher, the result of
partial covering of the continuum (e.g., Arav \etal\ 1999) or scattered light
(e.g., Ogle \etal\ 1999) filling in what would otherwise be black, saturated
absorption troughs.

The X-ray regime has supported the idea that the column densities towards BAL
quasars are quite high.  Green \& Mathur (1996) argued more than a decade ago
that {\it ROSAT} non-detections indicated column densities of greater than $N_{H}
\sim 10^{22}\perc2$, and deeper observations of BAL quasars by later X-ray
telescopes indicate typical column densities of $N_{H} \sim 10^{23}\perc2$ as
well as objects with columns in excess of $10^{24}\perc2$ (recently tabulated
by Punsly 2006).

While progress has been made in understanding some properties of BAL outflows
(see, e.g., Gallagher \& Everett 2007), many aspects of their intrinsic nature
remain poorly constrained.  The relationship between the ultraviolet and X-ray
absorbing material is not known for certain.  The location, geometry, physical
state, and chemical abundance of the absorbing material are poorly constrained
and model dependent.  It is not yet known whether outflows are present in
every quasar, and why their properties vary so dramatically (although there is
a strong luminosity dependence, e.g., Ganguly \etal\ 2007).

X-ray investigations have provided some progress.  In the X-ray regime, deep
observations of individual BAL quasars with the {\it XMM-Newton} and {\it Chandra}
observatories have led to a better understanding of the absorbing material.
The general result seems to be that BAL quasars have underlying intrinsic
X-ray properties consistent with those of unabsorbed quasars, and the absorber
is complex requiring fitting with models featuring some combination of
ionization and/or partial covering \citep{gallagher:06}.  Recent studies
investigating BAL quasars versus radio-loudness favoured a geometric model to
describe the observed properties, but could not explain strong polar BAL
quasars or the deficit of FRII sources within BAL quasars \citep{shankar:08}.
The dependence of the physical nature of the BAL outflows on properties such
as radio-loudness or the absorber ionization state needs to be observationally
established to better understand these systems (see, e.g., Dai, Shankar, \&
Sivakoff 2012).

To date, nearly all of the BAL quasars with observed X-ray spectra beyond mere
detections have been optically bright, blue, and radio-quiet, displaying only
high-ionization BALs (HiBALs); three exceptions are the cloverleaf quasar with
low-ionization BALs (LoBALs), H1413+117 (e.g., Chartas \etal\ 2004), the LoBAL
quasar Mrk 231 \citep{gallagher:02a,braito:04}, and the radio-loud BAL quasar
PKS 1004+130 (e.g., Miller \etal\ 2006).  Similar fractions of radio-loud and
radio-quiet quasars display BALs quasars (e.g., Brotherton \etal\ 1998; Becker
\etal\ 2000, 2001; Hewett\& Foltz 2003; Shankar \etal\ 2008), although BALs
are only very rarely seen in the spectra of powerful FRII radio-loud quasars
\citep{gregg:06}.  LoBAL quasars are probably rarer than radio-loud BAL
quasars, but so-called LoBAL quasars are also often reddened (e.g., Becker
\etal\ 2000; Brotherton \etal\ 2001, Sprayberry \& Foltz 1992, DiPompeo
\etal\ 2012a), and the BAL troughs can effectively wipe out rest-frame
ultraviolet light (e.g., Hall \etal\ 2002), making their true frequency
difficult to determine accurately.

FIRST J155633.8$+$351758 ($z=1.5008\pm0.0007$), hereafter FIRST J1556$+$3517,
was originally discovered as a red stellar object associated with a radio
source \citep{becker:97}, and was identified as the first radio-loud BAL
quasar based on its observed properties.  We will discuss this classification
later.  Its spectrum is unusual, even for BAL quasars, displaying not only
absorption from low-ionization species like Mg II, but also metastable Fe II
species, garnering it the subclass of `FeLoBAL' quasar.  It is also one of
the most optically polarized BAL quasars known \citep{brotherton:97}, and is
reddened by A$_{V} \approx 1.6$ \citep{najita:00}.
%The redshift was measured by \citet{brotherton:97} 
Furthermore, based on its radio variability,
FIRST J1556$+$3517 can be identified as a BAL quasar seen close to jet
on \citep{ghosh:07}.  The combination of extreme properties makes this quasar
an interesting target to study at all wavelengths.

\citet{brotherton:05} detected FIRST J1556$+$3517 at X-ray energies as part of
an exploratory {\it Chandra} survey of radio-loud BAL quasars.  All the
quasars in the survey are X-ray faint compared to unabsorbed radio-loud
quasars of similar luminosity.  Previous studies by \citet{miller:09} have
confirmed that BAL quasars appear X-ray weak relative to their non-BAL quasar
counterparts.  Compared to other LoBAL quasars, however, FIRST J1556$+$3517 is
relatively X-ray bright (0.0077 counts s$^{-1}$ with {\it Chandra} Advanced
CCD Imaging Spectrometer S-array (ACIS-S) in
the $0.35$--$8\kev$ energy band), making it a good target for deeper
follow-up.

We report here the results of a new $60\ks$ ACIS-S observation of FIRST
J1556$+$3517 with the {\it Chandra X-ray Observatory} (Section
\ref{sec:observations}). We discuss the lack of long and short-term
variability in Section \ref{sec:variability}. We explore a variety of models
to fit the X-ray spectrum (Section \ref{sec:spectral_modeling}). Our new
observations, in conjunction with other information from literature, allow us
to comprehensively investigate the observed and intrinsic spectral energy
distribution (SED) of FIRST J1556$+$3517 for the first time (Section
\ref{sec:spectral_energy_distribution}).  Finally, we discuss our results in
the context of how this extreme quasar fits into our understanding of the
broader population of BAL quasars and summarize our conclusions (Section
\ref{sec:discussion}).  We assume a cosmology defined by $(\Omega_{0},
\Omega_{\Lambda}, h_{100}) = (0.30, 0.70, 0.70)$, where $h_{100} = H_{0} /
100\hubble$.  Unless otherwise noted, error bars are $1\sigma$, and power-law
slopes ($\alpha$) are defined by the equation $F_{\nu} \propto \nu^{\alpha}$.

\section{OBSERVATIONS AND DATA REDUCTION}
\label{sec:observations}

We started the observations of our target on 2006 June 2 at 09:25:43 GMT (MJD:
53888.39287) with the {\it Chandra X-ray Observatory} ACIS-S3 in very faint
(VFAINT) mode.  We measured a total of 531 photons from the program object in
the $0.5$--$10\kev$ energy band over the $60\ks$ exposure time for a photon
rate of $8.85\pm0.38\times10^{-3}$ photons s$^{-1}$.  Fig.
\ref{fig:light_curve} shows the counts per second received versus the exposure
time for the 0.3--2, 2--10, and 0.3--10\kev\ energy bands.  The data were
initially processed using the standard {\it Chandra} X-ray Center pipeline
software.  Only the level 1 events file was used.  Additional processing was
carried out with the \verb+acis_process_events+ procedure of the CIAO 3.4
software.  Additional processing included removal of pixel randomization, and
selection of good Advanced Satellite for Cosmology and Astrophysics ({\it
  ASCA}) grades (0,2,3,4,6) and good status (ignoring the bits indicating
afterglow events).  The background light curve was inspected for temporal
fluctuations.  Fortunately none of the $60\ks$ exposure time was lost to
flaring.  The VFAINT 5$\times$5 event island was used to improve the filtering
of the background for cosmic rays.  From the calculated photon rate, the
estimated photon pile-up is $<\!\!1$\%.  As a precautionary measure, ``bad''
events filtered by the above-mentioned reduction procedure were visually
inspected to see if any of the source X-rays were rejected.  None was found.

The source photons were extracted by the \verb+psextract+ procedure from a
circular aperture with a $5''$ radius centred on the source.  The background
spectrum was taken from a concentric annulus void of any visible emission
sources or deficits with an inner radius of $10''$ and outer radius of $20''$.
The redistribution matrix (rmf) and auxiliary response file (arf) were
constructed using the standard procedure in CIAO 3.4.  For further spectral
analysis in Xspec (v.
12.3.1)\footnote{http://heasarc.gsfc.nasa.gov/docs/xanadu/xspec/index.html},
which is a detector independent, X-ray spectral-fitting program, the rmf and
arf energy grids were matched.  Energy ranges were restricted to the
0.5--10\kev\ range as below $0.5\kev$ the calibration is uncertain.  Note that
above $8.0\kev$ the effective area drops steeply and the particle background
increases, but for our analyses restricting the energy range to
0.5--8\kev\ had no effect on our conclusions.

\section{VARIABILITY}
\label{sec:variability}

\begin{figure}
\begin{center}
%\leavevmode
\includegraphics[width=2.5in]{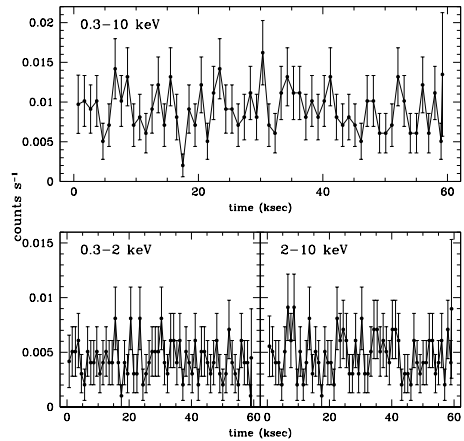}
\caption{X-ray light curve for FIRST J1556$+$3517.  Counts are binned in
  $\sim\!\!1\ks$ intervals, and energy bands are labelled in the top left
  corner of each panel.  All error bars are $1\sigma$ error bars as estimated
  by $\sqrt{n}$.}
\label{fig:light_curve}
\end{center}
\end{figure}

A visual inspection of the X-ray light curves (Fig. \ref{fig:light_curve})
did not reveal any significant short-term variability.  To quantify any
possible sub-$60\ks$ variability, we applied a Kolmogorov-Smirnov (KS) test to
the temporal cumulative photon count in the soft (0.3--2\kev), hard (2--10\kev),
and total energy (0.3--10\kev) energy bandpass.  We found the photon rate to be
consistent with a constant photon flux for the soft, hard and total energy
bandpasses at the $>\!\!90$\% level.

\citet{brotherton:05} estimated the photon rate to be $7.7\pm1.3\times10^{-3}$
photons s$^{-1}$ in the $0.35$--$8\kev$ energy band from a 5\ks\ {\it Chandra}
observation which started on 2000 May 20 at 10:51:30 GMT (MJD: 51684.45243).
They calculated an unabsorbed $0.35$--$8\kev$ flux of $6.5\times10^{-14}\flux$
using PIMMS\footnote{http://cxc.harvard.edu/toolkit/pimms.jsp.} with a photon
index $\Gamma=1.7$, and assuming a Galactic column density of
$2.0\times10^{20}\perc2$.  However, the most appropriate effective area for
cycle 1 was not employed in their flux calculation.  We have carried out a
better flux estimate using a more recent version of PIMMS with an effective
area more suitable (cycle 3) for the \citet{brotherton:05} observations.  Our
improved flux estimation using the same assumptions mentioned previously is
$5.1\pm0.9\times10^{-14}\flux$.  In this work, we also employed PIMMS with a
suitable effective area and the same assumptions of the \citet{brotherton:05}
observations for consistency.  Our new observation indicates a constant count
rate of $8.9\pm0.4\times10^{-3}$ photons s$^{-1}$ over the $60\ks$ exposure in
the $0.5$--$10\kev$ energy band, and any intrinsic variability is negligible
compared to the photon statistics.  Our measurement corresponds to an
unabsorbed flux of $7.1\pm0.3\times10^{-14}\flux$ in the $0.35$--$8\kev$
energy range using the same assumptions of $\Gamma=1.7$ and Galactic column.
This indicates a photon arrival rate that is higher than the previous epoch at
a $2.2\sigma$ level, only a marginal difference.  Evidence for long-term X-ray
variability is not conclusive.

We also note that FIRST J1556$+$3517 was observed a total of four times with
the {\it XMM-Newton} observatory.  All observations were inspected and
suffered from flaring events.  The portions of the observations suitable for
data extraction were $\ltorder20$\% of the total observation time, and did not
place further constraints on either the long-term or short-term X-ray
variability.

\section{EXPLAINING THE X-RAY SPECTRUM OF FIRST J1556$+$3517}
\label{sec:spectral_modeling}

There exist several explanations for the X-ray properties of FIRST
J1556$+$3517, as discussed by \citet{brotherton:05}.  They rejected the
simplest idea that FIRST J1556$+$3517 is an intrinsically normal quasar seen
through a large column density of neutral hydrogen.  \citet{brotherton:05}
used the radio-X-ray correlation \citep{brinkmann:00} to estimate the
intrinsic X-ray flux, and, comparing that result to the observed {\it Chandra}
count rate, determined that the observed X-rays were suppressed by a factor of
$49$.  This reduction in X-rays, if attributed to a neutral hydrogen absorber,
requires a column density of $6.0\times10^{23}\perc2$.  This large of a
column, however, would result in an extremely large hardness ratio (HR).  The
HR is defined as follows:
\begin{equation}
{\rm HR} = \frac{H-S}{H+S},
\end{equation}
where $S$ and $H$ are the total photon count in the soft band
($0.35$--$2\kev$) and the hard band ($2$--$8\kev$), respectively.  Brotherton
\etal\ concluded that the observed HR of $-0.1\pm0.2$ is incompatible with
that expected from a normal quasar absorbed by neutral hydrogen with column
density $\sim\!\!10^{23}\perc2$.

They preferred more complex explanations that reduce the observed X-ray flux
and not resulting in an HR that is inconsistent with observations.
These include: emission from an unobscured jet (mentioned below), an ionized
absorber, a partially covering neutral absorber or scattering/reflection.
Now, with a spectrum with more than 500 counts we can revisit the proposed
explanations in more detail.  We start with simple models and proceed to test
the more complex alternatives.

The best-fitting parameters are determined by minimizing the sum of the
squares of the deviations ($\chi^{2}$) with Marquardt-Levenberg optimization.
We restrict our fitting range to $0.5$--$10\kev$, and assume a Galactic column
density of $2.0\times10^{20}\perc2$ for all models \citep{dickey:90}, with
relative abundances defined by \citet{anders:89} and the cross-sections of
\citet{morrison:83}.  A study of the brightness temperature and radio
variability (time-scale of $\sim$1 year) of FIRST J1556$+$3517 by
\citet{ghosh:07} supports the presence of a beamed radio source.  Because of
this likelihood, the use of the radio-X-ray correlation to estimate the
intrinsic X-ray flux is suspect and therefore we prefer to permit
normalizations to vary freely.

\subsection{Power-law and neutral absorber models} 
\label{sub:power-law_plus_neutral_absorber}

\begin{figure}
\begin{center}
%\leavevmode
\includegraphics[width=2.5in]{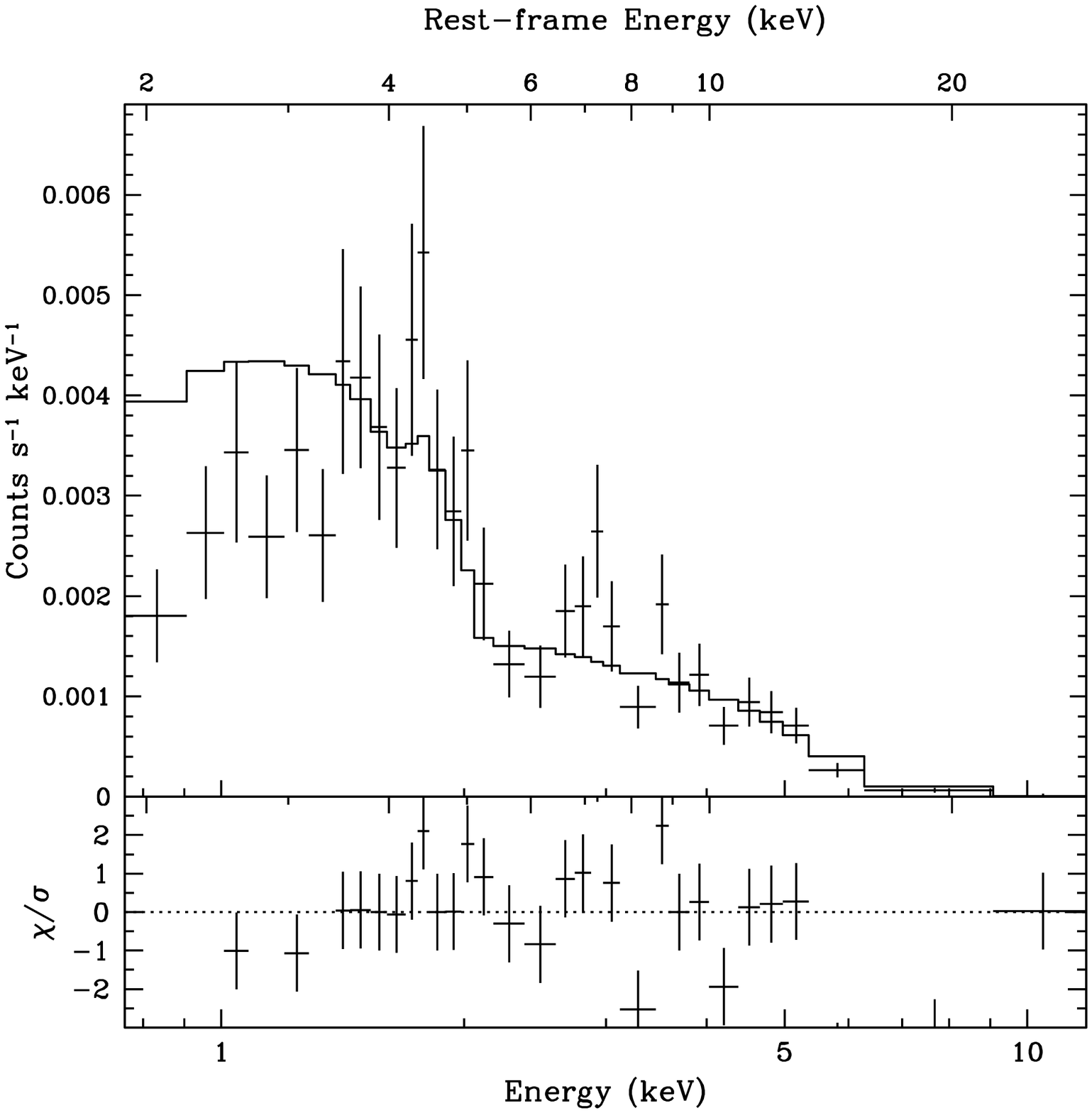}
\caption{The best-fitting power-law model for the observed energies
  $2$--$5\kev$.  The model includes a Galactic column density of
  $2.0\times10^{20}\perc2$.  The lower panel shows the normalized deviations
  ($\chi/\sigma$) to the of the bin values from the best-fitting model
  normalized to observational error bars ($\sigma$).  The panel range is
  selected to show the deviations of the points within the $2$--$5\kev$ energy
  range.  The missing values reflect deviations outside the range of the
  panel.  This model consistently overpredicts the X-ray photon flux for
  energies $\ltorder\!\!1.3\kev$.  Observed energies are given along the
  bottom X-axis, and rest-frame energies are given along the top axis of both
  panels.}
\label{fig:power_law}
\end{center}
\end{figure}

We start with a very simple model to confirm the presence of absorption.  We
fit a power-law model plus only Galactic extinction.  The fit quality was poor
(with a reduced $\chi^2$ or $\chi^{2}/\nu = 1.53$ where $\nu$ is the number of
degrees of freedom), and overpredicted the photon flux for energies
$<\!\!1.5\kev$.  Model fits are classified as statistically significant fits for
reduced $\chi^2 \ltorder 1.0$.  Henceforth, the quality of a model fit will be
reported in terms of their reduced $\chi^2$ unless otherwise noted.  The
best-fitting model resulted in a flat photon index of
$\Gamma=0.7^{+0.05}_{-0.06}$ where $\Gamma$ is defined by $N(E) = K
E^{-\Gamma}$ photons s$^{-1}$\perc2\ keV$^{-1}$.

To highlight the presence of absorption, we fit a power-law model only to the
observed frame $2$--$5\kev$ energy band (Fig. \ref{fig:power_law}).  The
resulting fit favoured a slightly softer photon index of
$\Gamma=0.85^{0.38}_{-0.23}$, but is no better than previous fit
($\chi^{2}/\nu = 1.36$).  Fig. \ref{fig:power_law} shows the inability of
the power-law model to accurately fit the $\ltorder1.5\kev$ energy range,
where the soft photons fall far below.  This result supports the conclusion of
\citet{brotherton:05} that the X-rays suffer absorption.

\begin{figure}
\begin{center}
%\leavevmode
\includegraphics[width=2.5in]{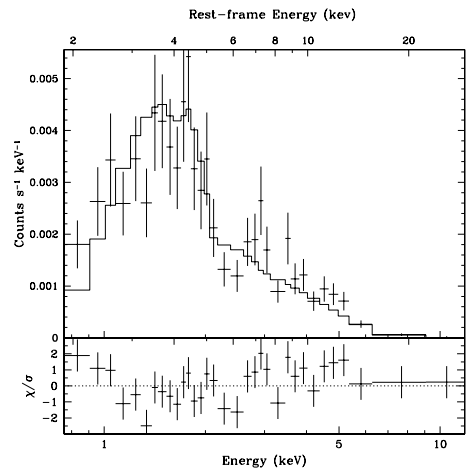}
\caption{The best-fitting neutral absorber model.  The model includes a
  Galactic column density of $2.0\times10^{20}\perc2$.  The lower panel shows
  the normalized deviations ($\chi/\sigma$) to the of the bin values from the
  best-fitting model normalized to observational error bars ($\sigma$).  The
  panel range is selected to show the deviations of the points within the
  $.8$--$10\kev$ energy range.  The missing values reflect deviations outside
  the range of the panel.  This model fails to accurately predict the X-ray
  photon flux for energies $\ltorder\!\!1\kev$.  Observed energies are given
  along the bottom X-axis, and rest-frame energies are given along the top
  axis of both panels.}
\label{fig:neutral_absorber}
\end{center}
\end{figure}

Next, we fixed the power-law photon index at $\Gamma=1.7$ from the average
photon index of radio-loud quasars \citep{brotherton:05}, but added a neutral
hydrogen column absorber at the quasar redshift.  Fig.
\ref{fig:neutral_absorber} shows the best fitting neutral absorber model with a
the photon index fixed at $\Gamma=1.7$.  We note that assuming a typical,
softer radio-quiet quasar photon index would be even more problematic than
what we find for the radio-loud case.  Allowing both the intrinsic neutral
absorber column density and the normalization to vary freely, the resulting
best-fitting model is only marginally better ($\chi^{2}/\nu = 1.32$), failing
to fit in particular the lowest energies.  If we let the intrinsic photon
index vary freely, then the best-fitting model is improved ($\chi^{2}/\nu =
0.98$).  This best-fitting model is statistically as significant as some of
the more complex models we investigate below, but requires an extremely flat,
and perhaps unrealistic, photon index of $\Gamma=1.2\pm0.1$.  The parameters
describing both fits are shown in Table \ref{tab:model_fits}.

%$\chi^{2}/\nu = 40.8/31$ $\chi^{2}/\nu = 29.3/30$

\subsection{Partially covering neutral absorber}
\label{sub:partially_covering_neutral_absorber}

\begin{figure}
\begin{center}
%\leavevmode
\includegraphics[width=2.5in]{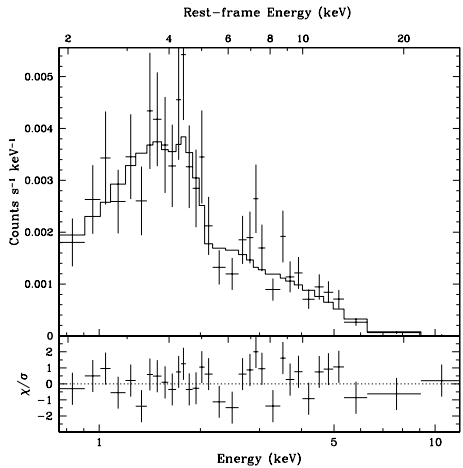}
\caption{The best-fitting partially covering neutral absorber model.  The
  intrinsic X-ray spectrum is assumed to be a power law with a photon index of
  $\Gamma=1.4\pm0.2$.  The parameters describing the best-fitting model are given
  in Table \ref{tab:model_fits}, model 2.  The bottom panel and panel axes for
  both panels are the same as Fig. \ref{fig:power_law}.  }
\label{fig:partially_covering_neutral_absorber}
\end{center}
\end{figure}

The neutral absorber model was unable to reproduce the observed soft X-ray
photons in the observed spectrum with a realistically steep intrinsic photon
index (e.g., $\Gamma=1.7$).  By allowing the neutral absorber to cover only a
fraction of the emitted X-ray spectrum, the uncovered fraction of the X-ray
emitter allows some soft photons to reach the observer unimpeded.  By varying
the covering fraction and the absorber column density, both the total X-ray flux
and HR may both be adjusted to fit the spectrum.  The partial
covering neutral absorber is modelled by the \verb+zpcfabs+ model.  Table 1
summarizes our best-fitting models, both fixing $\Gamma=1.7$ as well as
letting it vary, with a similar or better $\chi^{2}$ as the partially ionized
absorber models.  The total column densities for these models is of the order
of $10^{23}\perc2$ with covering fractions of $0.87$ and $0.81$, respectively.
The best-fitting partially covering neutral absorber model for a best-fitting
photon index $\Gamma=1.4\pm0.2$ is shown in Fig.
\ref{fig:partially_covering_neutral_absorber}.

\subsection{Partially ionized absorbing models}
\label{sub:partially_ionized_absorbing_models}

Partially ionized absorbers are capable of significantly reducing X-ray flux
without creating an excessively hard source.  To test such models, we applied
the ionized absorber model, \verb+absori+, provided in the XSPEC package.

\begin{figure}
\begin{center}
%\leavevmode
\includegraphics[width=2.5in]{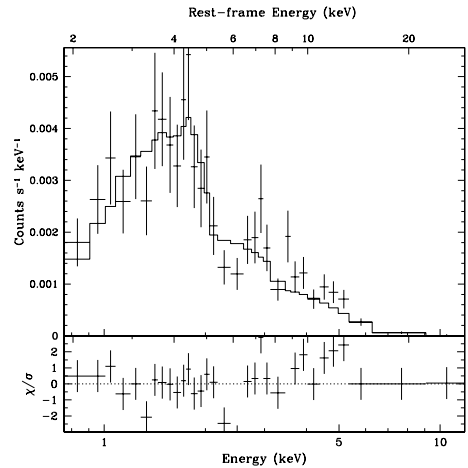}
\caption{The best-fitting ionized absorber model.  The intrinsic X-ray
  spectrum is assumed to be a power law with a photon index of $\Gamma=1.7$.
  The parameters describing the best-fitting model are given in Table
  \ref{tab:model_fits}.  The bottom panel is the same as the bottom panel of
  Fig. \ref{fig:power_law}.  Observed energies are given along the bottom
  X-axis.  Rest-frame energies are given along the top axis of both panels.}
\label{fig:ionized_absorber}
\end{center}
\end{figure}

We restricted the relative solar iron abundance to unity and held the gas
temperature fixed at $\sim\!\!3\times10^{4}\kelvin$.  Relative elemental
abundances are defined by \citet{anders:89}.  The ionizing photon index was
set at $\Gamma=1.7$ initially.  The absorber ionization state parameter $\xi$
as defined by \citet{done:92} was allowed to vary freely.  Typical values for
our fits were $\xi \gtorder 500.0$ ergs cm s$^{-1}$.  See Table
\ref{tab:model_fits} for exact values.  Absorber ionization state is defined
by the ionization parameter $\xi = L/nr^{2}$ of \citet{done:92} where L is the
integrated incident luminosity from $5\ev$ to $300\kev$, and $r$ is the
distance of the absorbing material of density $n$ from the illuminating
source.  Fig. \ref{fig:ionized_absorber} shows the best-fitting model
($\chi^{2}/\nu = 1.08$) to the observed X-ray spectrum.  For these parameters,
the column density for the ionized material required to match the observed
spectrum is $3.7\times10^{23}\perc2$.  We also fit a model using a fixed
photon index of $\Gamma=2.0$, similarly given in Table 1, which was worse
($\chi^{2}/\nu = 1.39$).
%($\chi^{2}/\nu = 32.4/30$) $\chi^{2}/\nu = 41.8/30$

To investigate the significance of the fit for the partially ionized absorber
model over the neutral absorber model ($\xi=0.0$ ergs cm s$^{-1}$), we test
for the significance that the ionization parameter ($\xi$) is greater than 0
with the photon index $\Gamma$ as a free parameter.  Our best-fitting models
favoured values of $\Gamma\approx1.3$, and values $\xi\approx300$ ergs cm
s$^{-1}$.  The partially ionized absorber model marginally improves the fit
over a neutral absorber model ($\sim$2$\sigma$) when photon indices are held
fixed at values consistent with quasar observations \citep{green:09}.

\subsection{Scattering/reflection}
\label{sub:scattering/reflection}

\citet{brotherton:97} showed that the polarization fraction of the optical
continuum for FIRST J1556$+$3517 is $\gtorder$ 13\%, and argued that the
polarization mechanism was scattering either by dust or hot electrons.  If the
polarization mechanism is scattering by electrons, then both the optical and
X-ray emission will be similarly scattered, and the minimum amount of the
intrinsic X-ray flux scattered into the line of sight is expected to be
$\sim$13\%.  This would imply that the intrinsic X-ray flux is no greater than
approximately 8 times the observed X-ray flux.

\citet{brotherton:05} argued for a ratio of $\sim$50 between the intrinsic and
observed X-ray flux based on the radio--X-ray correlation, much larger than
the factor of 8 determined from electron scattering.  They thus concluded that
electron scattering was not the mechanism in operation, although it is also
possible that the optical and X-ray geometries differ.

%-*-LaTeX-*-
%
% Last altered by $Author: rberring $ on $Date: 2007/11/07 23:15:57 $.  
%
% $Id: tab1.tex,v 1.9 2007/11/07 23:15:57 rberring Exp $
%
\begin{table}
%\tablecolumns{4}
%\tablewidth{0pt}
\caption{Model parameters}\label{tab:model_fits}
%\tablecaption{Model parameters for the best fit models.  All models include a
%  Galactic column density of $2.0\times10^{20}\perc2$.  All normalization
%  parameters have the units of photons s$^{-1}$ cm$^{-2}$ keV$^{-1}$ at
%  $1\kev$.  }
\begin{tabular}{rcll}
  \multicolumn{2}{c}{Property$^{\dag}$} &
  Model 1 & Model 2 \\ \hline

\multicolumn{4}{c}{Neutral Absorber} \\ \hline
N$_{\rm H}$ ($10^{22}\perc2$) & $=$ & {\bf $7.0_{-0.8}^{+1.0}$} 
                                    & {\bf $3.6_{-0.9}^{+1.1}$} \\
norm & $=$ & $2.2\pm0.2\times10^{-5}$ 
           & $1.2_{-0.2}^{+0.2}\times10^{-5}$ \\
$\Gamma$ & = & $1.7$ 
             & $1.2\pm0.1$ \\
($\chi^{2}/\nu$) & $=$ & {\bf $1.32~(40.9/31)$} 
                       & {\bf $0.98~(29.4/30)$} \\ \hline
\multicolumn{4}{c}{Ionized Absorber} \\ \hline

N$_{\rm H}$ ($10^{22}\perc2$) & $=$ & {\bf $27.2_{-8.5}^{+9.9}$} 
                                    & {\bf $37.5\pm10.8$}\\
$\xi$ (ergs cm s$^{-1}$) & $=$ & {\bf $620_{-363}^{+357}$} 
                               & {\bf $1056_{-573}^{+669}$} \\
$T_{g}$ (K) & $=$ & $3\times10^{4}$ 
                  & $3\times10^{4}$ \\
$\frac{[{\rm Fe}/{\rm H}]}{[{\rm Fe}/{\rm H}]_{\odot}}$ & $=$ & $1.0$ 
                                                            & $1.0$ \\
norm & $=$ & $1.2\pm0.1\times10^{-4}$ 
           & $2.3\pm0.2\times10^{-4}$\\
$\Gamma$ & = & $1.7$ 
             & $2.0$\\
($\chi^{2}/\nu$) & $=$ & {\bf $1.07~(32.2/30)$} 
                       & {\bf $1.39~(41.8/30)$} \\ \hline
\multicolumn{4}{c}{Partially Covering Neutral Absorber} \\ \hline

N$_{\rm H}$ ($10^{22}\perc2$) & $=$ & {\bf $14_{-2.5}^{+3.0}$} 
                                    & {\bf $9.5_{-4.1}^{+4.2}$} \\
covering fraction & $=$ & $0.87\pm0.03$ 
                        & $0.81_{-0.08}^{+0.07}$ \\
norm & $=$ & {\bf $1.3\pm0.1\times10^{-4}$} 
           & {\bf $6.0_{-2.4}^{+4.1}\times10^{-5}$} \\
$\Gamma$ & $=$ & $1.7$ 
               & $1.4\pm0.2$ \\
($\chi^{2}/\nu$) & $=$ & {\bf $0.98~(29.3/30)$} 
                       & {\bf $0.94~(27.3/29)$} \\ \hline
\end{tabular}

\medskip
{\em \dag} All values are quoted for the rest frame.  All fits include a
Galactic column density of $(N_{H} = 2.0\times10^{20}\perc2)$.  Errors
included with each property are $1\sigma$ errors.  Properties without errors
were held fixed with model 1 having a photon index of value 1.7, and model 2
with alternate or varying photon index values.
\end{table}

We note that in principle scattering need not appear differently in an
unresolved source from partial covering.  Some fraction of the light passes
through or around an absorber.  The partially covering neutral absorber models
of the previous section result in acceptable fits, and are consistent with the
$\sim$13\% polarization level.  That is, in the case where the polarization
efficiency is 100\%, both the optical and X-ray results may be explained by
the same geometry.  This would require that the intrinsic X-ray flux levels
found for the partial covering models be plausible, which they are (see
Section \ref{sec:spectral_energy_distribution}).

What about reflection from colder material?  In general, the presence of an Fe
K$\alpha$ line in a reflected X-ray spectrum depends on the ionization state
of the reflecting material.  With a possible rest-frame energy of $6.4\kev$
(neutral), $6.7\kev$ (He-like), and $6.96\kev$ (H-like), and an observed
redshift of $z=1.5008\pm0.0007$, we expect the Fe K$\alpha$ line to be
observed at $2.6\kev$ (neutral) , $2.7\kev$ (He-like), and $2.8\kev$ (H-like),
respectively.  Figs
\ref{fig:power_law}--\ref{fig:partially_covering_neutral_absorber} show a
possible emission feature centred at $\sim$2.9\kev, which has a rest-frame
energy of $7.2\kev$.  If the observed feature is an Fe K$\alpha$ line it must
be associated with outflowing material (at $>99\%$ confidence).  The possible
calculated outflow velocities are $5.0\pm0.6\times10^{4}\kms$ for a neutral Fe
K$\alpha$, $2.7\pm0.6\times10^{4}\kms$ for He-like, and
$1.6\pm0.6\times10^{4}\kms$ for H-like.  It is difficult to associate the
likely candidates with the possible emission feature observed at
$\sim$2.9\kev.  This is not very compelling evidence for the existence of a
reflected component.  Higher quality observations are required to resolve this
issue.

\subsection{The possibility of jet X-rays}
\label{sub:the_possibility_of_jet_x-rays} 

Alternatively, X-ray photons may be produced in the jet through synchrotron
and/or inverse-Compton processes.  Whether or not FIRST J1556$+$3517 is
intrinsically radio-loud or radio-quiet, it is clearly associated with a
compact, flat spectrum radio source that may signify a jet
\citep{reynolds:13}.  However, we argue that the origin of X-rays from FIRST
J1556$+$3517 are mostly of non-jet origin because of the following:

(i) Electrons that produce synchrotron emission at X-ray energies in the
presence of an $\sim$1 G magnetic field \citep[a reasonable assumption for
  $\la$pc-scale jets of quasars, e.g.,][]{ghisellini:10} have Lorentz factors
($\gamma$) of $\sim10^6$ with characteristic variability time-scales for
high-energy electrons radiating at X-ray energies from approximately minutes
to hours.  This variability is supported by numerous observations of blazars
\citep[e.g.,][]{takahashi:96,chiappetti:99,fossati:00,kataoka:00,edelson:01}.
If there were a significant contribution from a jet, we would expected to
observe variability during the $60\ks$ window shown in
Fig.\ \ref{fig:light_curve}.  But none is observed.

If the synchrotron emission component does extend to the X-ray band, it is
usually expected that the synchrotron peak is at or below the X-ray
frequencies \citep[e.g.,][]{landt:08}. In that case, the X-ray spectrum should
be steep ($\Gamma > 2$) in contrast to the observation.  While a jet may
contribute to the X-ray emission seen in FIRST J1556$+$3517, we find the
flatter X-ray spectrum along with the lack of short-term variability is
compelling evidence that the X-ray emission is not dominated by synchrotron
emission from a jet.

(ii) The X-rays can also be produced by inverse-Compton scattering of seed
photons by the relativistic electrons in the jet.  The seed photons may be
from outside the jet, such as, thermal emission from the accretion disc or
line emission from the broad line region (`external Compton' or EC process),
but may also originate from the synchrotron photons produced in the jet
(`synchrotron self-Compton' or SSC process).  Electrons responsible for
generating SSC or EC X-rays are less energetic ($\gamma$ $\sim$10$^2$--10$^3$)
and may have longer variability time-scales.  Furthermore, the spectra of the
X-ray photons produced by the SSC and EC processes can show flatter spectral
slopes.  Despite the possibility of not observing variability within a
$60\ks$ window, we show that the observed X-ray flux in FIRST J1556$+$3517
is much larger than what is expected via SSC or EC processes in its pc-scale
jet, and therefore unlikely to be dominated by emission from a jet.

Radio variability (time-scale of $\sim$1 yr) and the correspondingly high
brightness temperature indicate the presence of a beamed jet aligned within
$14^{\circ}$ of the line of sight \citep{ghosh:07}.  Monte Carlo simulations
of the correlation between viewing angle and radio power-law spectral indices
for BAL quasars supports a similar viewing angle ($\sim16^{\circ}$) from the
aforementioned radio power-law slopes \citep{dipompeo:12b}.  If we assume that
the radio core has an angular size of $\sim\!\!1$ mas (Jiang \& Wang 2003
report an upper limit of 20 mas), the spectral index of optically thin
synchrotron emission $\alpha=0.75$, and the Doppler factor of four from
\citet{ghosh:07}, then the estimated SSC flux of
the jet associated with FIRST J1556$+$3517 (using equation 1 from
\citet{ghisellini:93}) is $\sim\!\!10^{-21}\flux$.  Where we have assumed
energetic isotropic electrons well described a power-law energy distribution
entrained within a tangled, homogeneous magnetic field, a spherically
symmetric moving jet (see \citet{ghisellini:93}), and an observed frequency
similar to the self-absorption frequency to account for the flat radio
spectrum of the core emission.  This flux value is many orders of magnitude
lower than the observed {\it Chandra} X-ray flux ($\sim\!\!10^{-14}\flux$).
Therefore, SSC emission associated with the jet cannot explain the observed
X-ray properties, and is unlikely to be the dominant source of X-ray emission.

In most quasar jets the external photon field energy density is within a
factor of $\leq$100 of energy density of the magnetic
field\citep[e.g.,][]{ghisellini:10,giommi:12}.  Hence, the EC radiation is
within a factor of $\leq$100 of the SSC emission.  However, as mentioned
above, the estimated SSC flux is seven orders of magnitude smaller than the
observed \textit{Chandra} X-ray flux.  Therefore, the EC X-rays will also be
much smaller (approximately $5$ orders of magnitude) than observed.

(iii) If there is a significant contribution of jet emission in the X-rays, it is
likely that the optical-IR (OIR) emission will also be dominated by jet
synchrotron
emission\citep[e.g.,][]{chatterjee:08,marscher:08,jorstad:10,marscher:10,agudo:11,bonning:12}.
In that case the OIR emission will be significantly polarized. However,
\citet{brotherton:97} showed that the emission lines observed in the optical
are polarized at the same level as the optical continuum which strongly
suggests that the observed polarization is a result of scattering.  Hence, a
synchrotron explanation for the polarization of the OIR emission is not
required and in fact not likely.

(iv) It is evident from the SED that the X-ray emission in FIRST J1556$+$3517 is
lower than what is expected from a non-BAL radio-quiet quasar with similar
optical/UV emission. This is consistent with our conclusion that the jet does
not contribute significantly to the X-ray band in this object.

Studies of the distribution of radio--X-ray power-law slopes ($\alpha_{\rm rx}$)
of AGN with known jets indicates that the cores of flat radio spectrum quasars
have an $\alpha_{\rm rx}$ centred on a value of $0.5$ \citep{marshall:05} for
non-simultaneous observations, and is distributed differently than the mean
value of $-0.9$ for the extended emission associated with the jet
\citep{marshall:05,sambruna:06}.  Our calculated intrinsic $\alpha_{\rm rx}=-0.7$
places FIRST J1556$+$3517 between the core and jet distributions,
possibly indicating that some of the observed X-ray spectrum has a jet origin.
However, jet models are well described by simple power law or power law with
neutral absorber models.  If we accept the neutral absorber model, then 
the best-fitting observed photon index $\Gamma$ is uncharacteristically flat for
radio-quiet quasars.  For typical $\Gamma$ values, the neutral absorber model
poorly describes the X-ray spectrum, and favours more complicated models not
typical of jet model fits \citep{sambruna:06}.

While a comparison of our observations to detailed observations of jets by
\citet{marshall:05,sambruna:06,sambruna:07} indicate that FIRST
J1556$+$3517 is consistent with the most extreme knots, and cannot rule out
the possibility that a jet may be a significant contributor to the observed
X-ray spectrum, we find it improbable, and find the arguments concerning the
lack of X-ray variability presented in Section \ref{sec:variability} and the
arguments presented above compelling.  

\subsection{Analysis summary}
\label{sub:analysis_summary}

We confirm the results found by \citet{brotherton:05} that an absorber is
present, but a simple neutral absorber alone with a photon index like that of
normal quasars cannot accurately reproduce the observed X-ray spectrum.  The
best-fitting neutral absorber model required a column density of
$> 6.8\times10^{22}\perc2$ if the photon index is fixed at $\Gamma = 1.7$.
The neutral absorber model requires an intrinsic X-ray slope that is extremely
flat ($\Gamma=1.2$) to achieve a good fit.  This photon index is approximately
a $2\sigma$ deviation from the mean photon index observed for bright X-ray
sources \citep{george:00,tozzi:06} placing it on the tail of the distribution
for photon indices and therefore unlikely.  We prefer the more complex models.

Of the other models explored, several gave reasonable fits to the observed
X-ray spectrum.  The first of these is the partially ionized absorber.  While
this model does not help to explain other features like the presence of
polarized light in the optical spectrum, it is plausible.  In this case there
is strong absorption with column densities $N_{H} \approx
4.0\times10^{23}\perc2$.

We prefer the partially covering neutral absorber model, as it provides a
similarly good fit, but may also have the following additional explanatory
power.  Partial covering is consistent with the presence of scattered light,
which also explains the high optical polarization.  The best fitting values
are smaller, but still high, with column densities of $N_{H} \approx
1.0\times10^{23}\perc2$ and a covering fraction of $\sim85$\%.  The inferred
intrinsic X-ray flux level will be considered in light of the total SED in the
next section.

\section{SPECTRAL ENERGY DISTRIBUTION}
\label{sec:spectral_energy_distribution}

\begin{figure}
\begin{center}
%\leavevmode 
\includegraphics[width=2.5in]{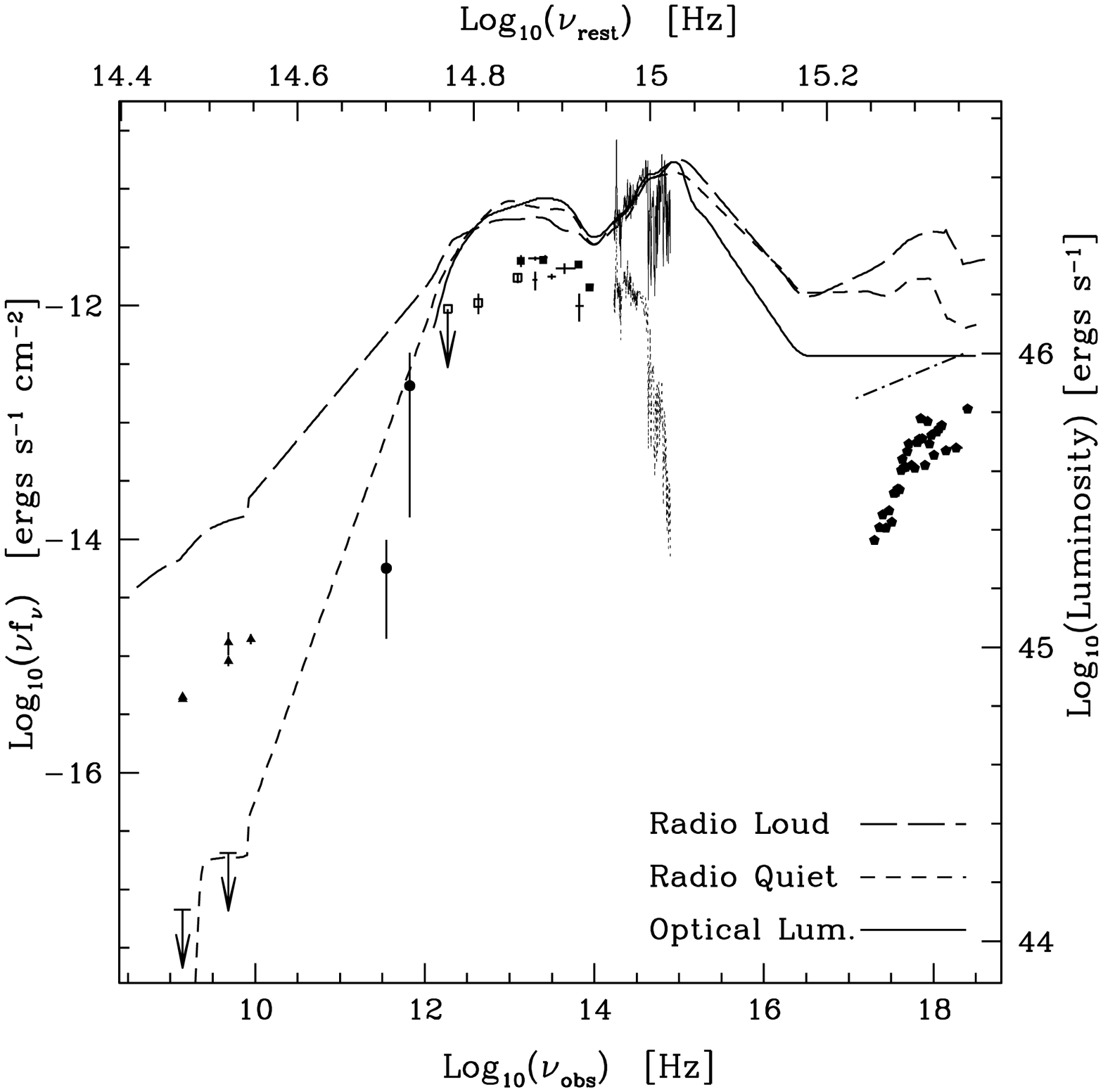}
\caption{SED for FIRST J1556$+$3517.  The optical luminous (solid) SED of
  \citet{richards:06}, and the radio-loud (long-dashed) and radio-quiet
  (dashed) \citet{elvis:94} SEDs are shown.  Data points are from the
  literature (for specific details see Section
  \ref{sec:spectral_energy_distribution}), and include both the reddened
  (dotted) and dereddened (thin solid) optical spectrum of \citet{najita:00}.
  The dereddened total-light spectrum assumes A$_{V} \approx 1.6$ and an SMC
  dust model.  Note the Elvis SEDs are approximately $7$ times brighter than
  the predicted intrinsic X-ray luminosity of the partially covering neutral
  absorber model (dot-dashed).  The SEDs are normalized to the dereddened
  optical spectrum, and all frequencies are observed frequencies.  All error
  bars shown are $1\sigma$ error bars, and the detection limit shown at
  $\log_{10}(\nu) \approx 12.2$ is a $3\sigma$ detection limit.  The two radio
  upper limits are the radio emission values corrected for beaming (see
  \citet{ghosh:07}).  The top and bottom axes give the base-10 logarithm of
  the photon frequency in the restframe and the observed frame, respectively.
  The left axis represents the observed flux, and the right axis represents
  the observed luminosity in units of $\log_{10}(\!\!\ergss)$.}
\label{fig:sed}
\end{center}
\end{figure}

With our {\it Chandra} observations, FIRST J1556$+$3517 has now been observed
across more than eight orders of magnitude in frequency, from radio
wavelengths through X-ray energies.  Fig. \ref{fig:sed} presents the SED.
X-ray observations on the right are from this paper, with individual points
representing flux bins as in previous figures.  Continuing left to lower
frequencies is the observed optical \citep[Keck spectrum;][]{brotherton:97},
the near-infrared \citep[Kitt Peak spectrum;][]{najita:00}, the mid-infrared
\citep[{\it Infrared Space Observatory};][]{clavel:98}, the far-infrared ({\it
  Spitzer}, Farrah \etal\ 2007, and {\it Wilkinson Microwave Anisotropy
  Probe}, DiPompeo \etal\ 2013), the millimetre \citep[Submillimetre
  Common-User Bolometer Array on James Clerk Maxwell Telescope;][]{lewis:03},
and the radio [Green Bank Telescope, Gregory \& Condon 1991; the VLA(Very
  Large Array) Faint Images of the Radio Sky at Twenty-cm (FIRST) Survey,
  Becker, White, \& Helfand 1995; and the VLA, DiPompeo \etal\ 2011].

In addition to the observed SED, we make corrections in three regimes to
present an estimate of the intrinsic SED.  In the X-ray regime, we show the
intrinsic unabsorbed power law of our preferred partially covered neutral
absorber model.  In the optical/near-infrared regime, we use the total light
spectrum from \citet{najita:00}, dereddened by 1.6 visual magnitudes of
extinction using an extinction law appropriate for the Small Magellanic Cloud (SMC),
which provides a final spectral index of $-0.5$, matching average quasar
spectra.  This is consistent, certainly at the level of our SED plotted here
on a log-log scale, with all the estimates of the reddening
\citep{brotherton:97,clavel:98,najita:00}, which are all $A_{V} \sim 1.6$.
Finally, in the radio regime, we show the correction for Doppler beaming based
on the lower limit to the Doppler factor of 4 from \citet{ghosh:07}
(assuming an intrinsic radio spectral index of $0$).  For reference, the average
radio-loud and radio-quiet quasar SEDs of \citet{elvis:94} representative of
lower luminosity quasars and the optically luminous quasar SED of
\citet{richards:06}, a better match to FIRST J1556$+$3517, are scaled to
match the dereddened optical flux and are also shown in Fig. \ref{fig:sed}.
We note that we in particular matched the dereddened spectrum in the
rest-frame optical where issues with polarization and BALs are minimized, and
that the slope matched well with those of the SEDs.

A few points need to be mentioned. First, \citet{zhou:06} and \citet{ghosh:07}
have shown that a class of BAL quasars based on radio source variability and
brightness temperature arguments appear to be consistent with close to pole-on
views, and hence polar outflows.  This conclusion is supported by earlier
arguments put forward by \citet{becker:00} for a variety of outflow
orientations of radio-selected BAL quasars based on a range of radio spectral
indices consistent with both polar and edge-on geometries.  More recent
studies by \citet{dipompeo:12a} compared the radio spectral indices and
viewing angles of BAL quasars with unabsorbed quasars.  Their analysis
confirms a large overlap in the viewing angle distribution of both samples
with both distributions extending to the jet axis supporting the existence of
BAL quasars with polar outflows.  The SED of FIRST J1556$+$3517 shows a flat
radio spectrum consistent with that of a pole-on source.  \citet{ghosh:07}
specifically identify FIRST J1556$+$3517 as close to pole-on, with a jet angle
to the line of sight less than $14^{\circ}$ and a minimum Doppler factor of
4.

These considerations are relevant to the classification of FIRST J1556$+$3517
as a radio-loud quasar, and in understanding its X-ray properties.
\citet{becker:97}, in fact, claimed this object as the first radio-loud BAL
quasar, which is based on its apparent observed properties.  One of the ways
to classify radio-loudness is by using $\log_{10}(R^{*})$, the ratio of
rest-frame 5 GHz to 2500 \AA\ flux, with unity separating the classes
\citep{stocke:92}.  Becker \etal\ reported $\log_{10}(R^{*}) > 3$, based on
the observed optical and radio data.  \citet{najita:00} revised the
$\log_{10}(R^{*})$ value to 0.9 based on dereddening the optical spectrum.  If
we additionally correct the radio data for beaming based on the Doppler factor
and assume a flat radio spectrum as observed, we find that $\log_{10}(R^{*}) <
-0.9$ and is consistent with radio-quiet quasars.  Radio-quiet quasars are not
radio silent, and have been observed to have relativistic jets and evidence of
beaming like this before \citep{falcke:96,blundell:03,wang:06}.  Making the
beaming correction places the upper limits on the \citet{elvis:94} radio-quiet
quasar SED.

\citet{gallagher:06} found LoBAL quasars to be optically reddened and more
deficient in observed X-rays than quasars showing only HiBALs,
and we see similar behaviour in FIRST J1556$+$3517; the target of our study is
reddened and is X-ray deficient with apparent absorbing column densities
$N_{H} > 10^{23}\percm2$ (from our preferred models shown in Table
\ref{tab:model_fits}).  However, a study by \citet{streblyanska:10} found that
LoBAL quasars have lower column densities ($N_{H} < 10^{22}\percm2$) than
HiBAL quasars.  Their conclusion may have been biased by the selection of
X-ray bright BAL quasars with relatively high S/N spectra suitable for X-ray
spectral analysis.

In the case of FIRST J1556$+$3517 we can directly determine the intrinsic
X-ray brightness relative to the optical.  After the corrections for
optical/UV dereddening and X-ray absorption for our favoured reddening values
and X-ray model, our intrinsic $\alpha_{\rm ox} = -1.7$.
%In this calculation straight forward assumptions regarding the
%determination of the intrinsic luminosities were made, and the formal errors
%in the calculation of $\alpha_{ox}$ from these assumptions is small, but the
%systematic error resulting from model assumptions is unquantifiable.  
We define $\alpha_{\rm ox}$ to be the spectral index of a power-law between the
monochromatic luminosity $L_{\nu}$ at the rest-frame optical $2500$\AA\ and
X-ray $2\kev$ (in $\monolum$), or
\begin{equation}
\alpha_{\rm ox} = \frac{\log_{10}[L_{\nu}(2500\;{\rm
\AA})]-\log_{10}[L_{\nu}(2\kev)]} {\log_{10}[\nu(2500\;{\rm
\AA})]-\log_{10}[\nu(2\kev)]}.
\end{equation}  
The intrinsic $\alpha_{\rm ox}$ value is clearly smaller than expected for either
the radio-loud or radio-quiet quasar SEDs of \citet{elvis:94}, by nearly an
order of magnitude, but those SEDs were constructed for lower luminosity
quasars.  Using a more up-to-date result for the dependence of $\alpha_{\rm ox}$
on luminosity for radio-quiet quasars \citep{steffen:06}, we calculate that
FIRST J1556$+$3517 should have $\alpha_{\rm ox} = -1.73 \pm 0.35$.
This is not far from our estimate and is consistent with our new observations
and the optically luminous SED of Richards \etal

Other apparently pole-on BAL quasars have been observed at X-ray energies.
A study by \citet{wang:08} of four randomly selected polar BAL quasars
using {\it XMM-Newton} detected two of them.  The four quasars were selected
randomly from a sample of eight BAL quasars pulled from the Sloan Digital Sky
Survey quasar catalogue \citep{york:00}.  The brightness temperatures were
determined from their large radio band variability and far exceeded the
inverse Compton limit ($10^{12}$ K).  This was taken as compelling evidence
for the presence of a relativistic beamed jet towards the observer
\citep{zhou:06}.  These two detections show no clear evidence for X-ray
absorption from neutral hydrogen, and the limit of one of the non-detections
is also consistent with no absorption.  The final non-detection, of FIRST
J210757$-$062010, an FeLoBAL quasar, is likely significantly absorbed. They
conclude, within the limits of their data, that any absorption if present must
be complex (e.g., an ionized or partially covering absorber) or that there may
be X-rays contributed from a radio jet outside the BAL region.

We can also consider the SEDs of quasars displaying BALs more generally, too.
\citet{gallagher:07b} examined the SEDs of 38 BAL quasars, primarily bright
blue radio-quiet BAL quasars from the Large Bright Quasar Survey (LBQS;
Hewett, Foltz, \& Chaffee 1995; 2001).  They noted that BAL quasars have
optical to mid-infrared fluxes similar to those of normal quasars, although
for a larger sample and more careful comparison, BAL quasars appear more
likely to have a small mid-infrared excess \citep{dipompeo:13}.
\citet{farrah:07} suggested that FeLoBAL quasars in particular have
far-infrared excesses characteristic of enhanced star formation.
\citet{lazarova:12} found inconclusive evidence that far-IR luminosities of
LoBAL quasars differ from non-LoBAL quasars, but did suggest that the
IR-luminous LoBAL quasars have pronounced star formation rates in comparison
to their non-LoBAL counterparts possibly implying a brief period during the
LoBAL phase that quenched star formation rates to normal non-LoBAL levels.
FIRST J1556$+$3517 appears to have an excess based on the observed data, but
after dereddening the optical and scaling, the mid-infrared is deficient
compared to the Elvis \etal\ and Richards \etal\ SEDs.

To summarize, we have assembled the observed SED of FIRST J1556$+$3517 and
made corrections to obtain the intrinsic SED.  We can now characterize this
FeLoBAL quasar as a quasar consistent with a polar outflow and a normal
intrinsic $\alpha_{\rm ox}$ for its luminosity.  The radio emission is consistent
with a beamed radio-quiet quasar.  The mid and far-infrared fluxes are below
average relative to the dereddened the optical emission, suggesting that the
dust covering fraction and the star formation rate are not enhanced relative
to the typical quasar.

\section{Conclusions}
\label{sec:discussion}

We can reach a few conclusions about FIRST J1556$+$3517: the X-rays suffer
absorption from an intervening column of the order of a few times
$10^{23}\perc2$ or less, the absorption is likely complex (e.g., partial
covering or partial ionization), and the intrinsic X-ray level is consistent
with that of radio-quiet quasars of similar optical luminosity.  These
conclusions are similar to those that have been reached for most other BAL
quasars (mostly HiBAL quasars) with deep X-ray observations.  The absorber,
while possessing a substantial column density, is not Compton-thick.

Of the satisfactory models we fit to our spectrum, we have a preference for
the neutral absorber with partial covering.  From the ultraviolet iron
absorption, we know that very low-ionization material is along the line of
sight.  We also know that there is significant scattered light present based
on the spectropolarimetry of \citet{brotherton:97}, and scattered light may be
interpreted as partial covering.  Now that we understand FIRST J1556$+$3517 is
a beamed luminous radio-quiet quasar, and that our determination of the
intrinsic X-rays is consistent with this classification, we can explain the
partial covering of the X-ray absorber and the optical polarization results
simultaneously with electron scattering.  Testing this idea will require
knowledge of the X-ray polarization.  Still, we make the suggestion that
partial covering of the X-ray source is present and represents a manifestation
of non-axisymmetric equatorial electron scattering around a polar outflow seen
at a small angle.
 
The polar-outflow BAL quasars show the full range of ionization, with HiBALs,
LoBALs, and FeLoBALs, and no one has yet identified any clear distinguishing
features from other BAL quasars aside from their radio properties.  Their
luminosities and BAL terminal velocities fall among those of other BAL quasars
\citep{ganguly:08}, within the envelope thought to indicate radiative
acceleration.  They may be differently driven, but there is as yet no solid
evidence to support that.  Outflow along the jet direction suggests other
possible acceleration mechanisms (e.g., jet entrainment).  \citet{wang:08}
suggested possible differing acceleration mechanisms for polar BAL quasars,
but more recent studies disagree \citep{dipompeo:12a}.

We have added FIRST J1556$+$3517 to the short but growing list of BAL quasars
with spectroscopic observation in the X-ray band.  In addition to the
aforementioned LoBAL quasars Mrk 231, H1413+117, and radio-loud PG 1004+130,
these other BAL quasars, all radio-quiet HiBAL quasars, include
PG 1411+442, PG 1535+547 and PG 2112+059 \citep{gallagher:02b}, PG 1115+080
\citep{chartas:03}, APM 08279+5255 \citep{chartas:02}, Q1246--057 and SBS
1542+541 \citep{grupe:03}, UM 425 \citep{aldcroft:03}, CSO 755, Q 0000--263,
and RX J0911.4+0551 \citep{page:05}.  We have also added FIRST J1556$+$3517 to
the shorter list of BAL quasars with high-quality SEDs all the way from radio
to X-rays.  These are a subset of the above, primarily the brightest and
lowest redshift sources, which includes the PG quasars, APM 08279+5255, and
Mrk 231 (see also Gallagher \etal\ 2007 for lower quality SEDs of 38 LBQS BAL
quasars).

%\acknowledgments 
\section*{ACKNOWLEDGEMENTS}

We acknowledge support from {\it Chandra} Award No. GO6-7105X and from the US
National Science Foundation through grant AST 05-07781 (MSB).  ZS acknowledge
support by NASA under grant NNG05GD03G issued through the Office of Space
Science and by National Natural Science Foundation of China through grant
10643001.  This work was performed under the auspices of the US Department
of Energy by the University of California, Lawrence Livermore National
Laboratory under Contract No. W-7405-Eng-48.


\begin{thebibliography}{}
\bibitem[\protect\citeauthoryear{Agudo \etal}{2011}]{agudo:11}Agudo, I., \etal\ 2011, \apj, 726, L13
\bibitem[\protect\citeauthoryear{Aldcroft \& Green}{2003}]{aldcroft:03} Aldcroft, T.~L., \& Green,
P.~J.\ 2003, \apj, 592, 710
\bibitem[\protect\citeauthoryear{Anders \& Grevesse}{1989}]{anders:89} Anders, E., \& Grevesse, N.\
1989, \gca, 53, 197
%\bibitem[\protect\citeauthoryear{Arav \etal}{1999}]{arav:99} Arav, N., Becker, R.~H.,
%Laurent-Muehleisen, S.~A., Gregg, M.~D., White, R.~L., Brotherton, M.~S., \&
%de Kool, M.\ 1999, \apj, 524, 566
\bibitem[\protect\citeauthoryear{Arav \etal}{2001}]{arav:01} Arav, N., Brotherton, M.~S., Becker,
R.~H., Gregg, M.~D., White, R.~L., Price, T., \& Hack, W.\ 2001, \apj, 546,
140
%\bibitem[\protect\citeauthoryear{Balucinska-Church \& McCammon}{1992}]{balucinska-church:92}
%Balucinska-Church, M., \& McCammon, D.\ 1992, \apj, 400, 699
\bibitem[\protect\citeauthoryear{Becker \etal}{1995}]{becker:95} Becker, R.~H., White, R.~L., \&
  Helfand, D.~J.\ 1995, \apj, 450, 559
\bibitem[\protect\citeauthoryear{Becker \etal}{1997}]{becker:97} Becker, R.~H., Gregg, M.~D., Hook,
I.~M., McMahon, R.~G., White, R.~L., \& Helfand, D.~J.\ 1997, \apjl, 479, L93
\bibitem[\protect\citeauthoryear{Becker \etal}{2000}]{becker:00} Becker, R.~H., White, R.~L., Gregg,
M.~D., Brotherton, M.~S., Laurent-Muehleisen, S.~A., \& Arav, N.\ 2000, \apj,
538, 72
\bibitem[\protect\citeauthoryear{Becker \etal}{2001}]{becker:01} Becker, R.~H., \etal\ 2001, \apjs,
135, 227
\bibitem[\protect\citeauthoryear{Bonning \etal}{2012}]{bonning:12}Bonning,
  E. W. \etal\ 2012, \apj, 756, 13
\bibitem[\protect\citeauthoryear{Blundell \etal}{2003}]{blundell:03} Blundell, K.~M., Beasley, A.~J.,
\& Bicknell, G.~V.\ 2003, \apjl, 591, L103
%\bibitem[\protect\citeauthoryear{Brinkmann \etal}{1999}]{brinkmann:99} Brinkmann, W., Wang, T.,
%Matsuoka, M., \& Yuan, W.\ 1999, \aap, 345, 43
\bibitem[\protect\citeauthoryear{Braito \etal}{2004}]{braito:04} Braito, V., \etal\ 2004, \aap, 420, 79
\bibitem[\protect\citeauthoryear{Brinkmann \etal}{2000}]{brinkmann:00} Brinkmann, W.,
Laurent-Muehleisen, S.~A., Voges, W., Siebert, J., Becker, R.~H., Brotherton,
M.~S., White, R.~L., \& Gregg, M.~D.\ 2000, \aap, 356, 445
\bibitem[\protect\citeauthoryear{Brotherton \etal}{2001}]{brotherton:01} Brotherton, M.~S., Arav, N.,
Becker, R.~H., Tran, H.~D., Gregg, M.~D., White, R.~L., Laurent-Muehleisen,
S.~A., \& Hack, W.\ 2001, \apj, 546, 134
%\bibitem[\protect\citeauthoryear{Brotherton \etal}{2006}]{brotherton:06} Brotherton, M.~S., De Breuck,
%C., \& Schaefer, J.~J.\ 2006, \mnras, 372, L58
\bibitem[\protect\citeauthoryear{Brotherton \etal}{2005}]{brotherton:05} Brotherton, M.~S.,
Laurent-Muehleisen, S.~A., Becker, R.~H., Gregg, M.~D., Telis, G., White,
R.~L., \& Shang, Z.\ 2005, \aj, 130, 2006
\bibitem[\protect\citeauthoryear{Brotherton \etal}{1997}]{brotherton:97} Brotherton, M.~S., Tran,
H.~D., van Breugel, W., Dey, A., \& Antonucci, R.\ 1997, \apjl, 487, L113
%\bibitem[\protect\citeauthoryear{Brotherton \etal}{2002}]{brotherton:02} Brotherton, M.~S., Croom,
%S.~M., De Breuck, C., Becker, R.~H., \& Gregg, M.~D.\ 2002, \aj, 124, 2575
\bibitem[\protect\citeauthoryear{Brotherton \etal}{1998}]{brotherton:98} Brotherton, M.~S., van
Breugel, W., Smith, R.~J., Boyle, B.~J., Shanks, T., Croom, S.~M., Miller, L.,
\& Becker, R.~H.\ 1998, \apjl, 505, L7
\bibitem[\protect\citeauthoryear{Chartas \etal}{2003}]{chartas:03} Chartas, G., Brandt, W.~N., \&
Gallagher, S.~C.\ 2003, \apj, 595, 85
\bibitem[\protect\citeauthoryear{Chartas \etal}{2002}]{chartas:02} Chartas, G., Brandt, W.~N.,
Gallagher, S.~C., \& Garmire, G.~P.\ 2002, \apj, 579, 169
\bibitem[\protect\citeauthoryear{Chartas \etal}{2004}]{chartas:04} Chartas, G., Eracleous, M., Agol,
E., \& Gallagher, S.~C.\ 2004, \apj, 606, 78
\bibitem[\protect\citeauthoryear{Chatterjee \etal}{2008}]{chatterjee:08} Chatterjee, R. \etal\ 2008, \apj, 689, 79 
\bibitem[\protect\citeauthoryear{Chiappetti \etal}{1999}]{chiappetti:99}
  Chiappetti, L., \etal\ 1999, \apj, 521, 552
\bibitem[\protect\citeauthoryear{Clavel}{1998}]{clavel:98} Clavel, J.\ 1998, \aap, 331, 853
\bibitem[\protect\citeauthoryear{Dai, Shankar, \& Sivakoff}{2012}]{dai:12} Dai
  X., Shankar F., Sivakoff G.~R., 2012, \apj, 757, 180
\bibitem[\protect\citeauthoryear{Dickey \& Lockman}{1990}]{dickey:90} Dickey, J.~M., \& Lockman, F.~J.\
1990, \araa, 28, 215
\bibitem[\protect\citeauthoryear{DiPompeo \etal}{2013}]{dipompeo:13} DiPompeo, M.~A., Runnoe, J.~C.,
  Brotherton, M.~S., \& Myers, A.~D.\ 2013, \apj, 762, 111
\bibitem[\protect\citeauthoryear{DiPompeo \etal}{2012a}]{dipompeo:12a} DiPompeo, M.~A., 
Brotherton, M.~S., Cales, S.~L., \& Runnoe, J.~C.\ 2012, \mnras, 427, 1135 
\bibitem[\protect\citeauthoryear{DiPompeo \etal}{2012b}]{dipompeo:12b} DiPompeo, M.~A., 
Brotherton, M.~S., \& De Breuck, C.\ 2012, \apj, 752, 6 
\bibitem[\protect\citeauthoryear{DiPompeo \etal}{2011}]{dipompeo:11} DiPompeo, M.~A., Brotherton,
  M.~S., De Breuck, C., \& Laurent-Muehleisen, S.\ 2011, \apj, 743, 71
\bibitem[\protect\citeauthoryear{Done \etal}{1992}]{done:92} Done, C., Mulchaey, J.~S., Mushotzky,
R.~F., \& Arnaud, K.~A.\ 1992, \apj, 395, 275
\bibitem[\protect\citeauthoryear{Edelson \etal}{2001}]{edelson:01} Edelson,
  R., Griffiths, G., Markowitz, A., Sembay, S., Turner, M. J. L., \& Warwick,
  R. 2001, \apj, 554, 274
\bibitem[\protect\citeauthoryear{Fossati \etal}{2000}]{fossati:00} Fossati,
  G., \etal\ 2000, \apj, 541, 153
\bibitem[\protect\citeauthoryear{Elvis \etal}{1994}]{elvis:94} Elvis, M., \etal\ 1994, \apjs, 95, 1
\bibitem[\protect\citeauthoryear{Farrah \etal}{2007}]{farrah:07} Farrah, D., Lacy, M., Priddey, R.,
  Borys, C., \& Afonso, J.\ 2007, \apjl, 662, L59
\bibitem[\protect\citeauthoryear{Falcke \etal}{1996}]{falcke:96} Falcke, H., Patnaik, A.~R., \&
Sherwood, W.\ 1996, \apjl, 473, L13
\bibitem[\protect\citeauthoryear{Gallagher \etal}{2002a}]{gallagher:02a} Gallagher, S.~C., Brandt,
W.~N., Chartas, G., \& Garmire, G.~P.\ 2002a, \apj, 567, 37
\bibitem[\protect\citeauthoryear{Gallagher \etal}{2002b}]{gallagher:02b} Gallagher, S.~C., Brandt,
W.~N., Chartas, G., Garmire, G.~P., \& Sambruna, R.~M.\ 2002b, \apj, 569, 655
\bibitem[\protect\citeauthoryear{Gallagher \etal}{2006}]{gallagher:06} Gallagher, S.~C., Brandt, W.~N.,
  Chartas, G., Priddey, R., Garmire, G.~P., \& Sambruna, R.~M.\ 2006, \apj,
  644, 709
\bibitem[\protect\citeauthoryear{Gallagher \& Everett}{2007}]{gallagher:07a} Gallagher, S.~C., \&
  Everett, J.~E.\ 2007, in Ho L. C., Wang J.-M., eds, ASP Conf. Ser. Vol 373,
  The Central Engine of Active Galactic Nuclei, Astron. Soc. Pac., San
  Francisco, p. 305
\bibitem[\protect\citeauthoryear{Gallagher \etal}{2007}]{gallagher:07b} Gallagher, S.~C., Hines, D.~C.,
Blaylock, M., Priddey, R.~S., Brandt, W.~N., \& Egami, E.~E.\ 2007, \apj, 665,
157
\bibitem[\protect\citeauthoryear{Ganguly \etal}{2007}]{ganguly:07} Ganguly, R., Brotherton, M.~S.,
Cales, S., Scoggins, B., Shang, Z., \& Vestergaard, M.\ 2007, \apj, 665, 990
\bibitem[\protect\citeauthoryear{Ganguly \& Brotherton}{2008}]{ganguly:08} Ganguly, R., \& Brotherton,
M.~S.  2008, \apj, 672, 102
\bibitem[\protect\citeauthoryear{George \etal}{2000}]{george:00} George, I.~M., Turner, T.~J., Yaqoob,
T., Netzer, H., Laor, A., Mushotzky, R.~F., Nandra, K., \& Takahashi, T.\
2000, \apj, 531, 52
\bibitem[\protect\citeauthoryear{Ghosh \& Punsly}{2007}]{ghosh:07} Ghosh, K.~K., \& Punsly, B.\ 2007,
  \apjl, 661, L139
\bibitem[\protect\citeauthoryear{Ghisellini \etal}{1993}]{ghisellini:93} Ghisellini, G., Padovani, P.,
Celotti, A., \& Maraschi, L.\ 1993, \apj, 407, 65
\bibitem[\protect\citeauthoryear{Ghisellini \etal}{2010}]{ghisellini:10} Ghisellini, G., Tavecchio, F.,
  Foschini, L., Ghirlanda G., Maraschi L., CelottiA., 2010, \mnras, 402, 497
\bibitem[\protect\citeauthoryear{Giommi \etal}{2012}]{giommi:12} Giommi, P.
  \etal\ 2012, \aap, 541, A160
%\bibitem[\protect\citeauthoryear{Green \etal}{2001}]{green:01} Green, P.~J., Aldcroft, T.~L., Mathur,
%S., Wilkes, B.~J., \& Elvis, M.\ 2001, \apj, 558, 109
\bibitem[Green \etal(2009)]{green:09} Green, P.~J. \etal\ 2009, \apj, 690, 644
%\bibitem[\protect\citeauthoryear{Gregg \etal}{2000}]{gregg:00} Gregg, M.~D., Becker, R.~H., Brotherton,
%M.~S., Laurent-Muehleisen, S.~A., Lacy, M., \& White, R.~L.\ 2000, \apj, 544,
%142
\bibitem[\protect\citeauthoryear{Gregg \etal}{2006}]{gregg:06} Gregg, M.~D., Becker, R.~H., \& de
Vries, W.\ 2006, \apj, 641, 210
\bibitem[\protect\citeauthoryear{Gregory \& Condon}{1991}]{gregory:91} Gregory, P.~C., \& Condon,
  J.~J.\ 1991, \apjs, 75, 1011
\bibitem[\protect\citeauthoryear{Grupe \etal}{2003}]{grupe:03} Grupe, D., Mathur, S., \& Elvis, M.\
2003, \aj, 126, 1159
%\bibitem[\protect\citeauthoryear{Hall \etal}{1997}]{hall:97} Hall, P.~B., Martini, P., Depoy, D.~L., \&
%Gatley, I.\ 1997, \apjl, 484, L17
\bibitem[\protect\citeauthoryear{Hamann \etal}{1993}]{hamann:93} Hamann, F., Korista, K.~T., \& Morris,
S.~L.\ 1993, \apj, 415, 541
\bibitem[\protect\citeauthoryear{Hewett \& Foltz}{2003}]{hewett:03} Hewett, P.~C., \& Foltz, C.~B.\
2003, \aj, 125, 1784
\bibitem[\protect\citeauthoryear{Hewett \etal}{1995}]{hewett:95} Hewett, P.~C., Foltz, C.~B., \&
Chaffee, F.~H.\ 1995, \aj, 109, 1498
\bibitem[\protect\citeauthoryear{Hewett \etal}{2001}]{hewett:01} Hewett, P.~C., Foltz, C.~B., \&
Chaffee, F.~H.\ 2001, \aj, 122, 518
\bibitem[\protect\citeauthoryear{Jiang \& Wang}{2003}]{jiang:03} Jiang, D.~R., \& Wang, T.~G.\ 2003,
\aap, 397, L13
\bibitem[\protect\citeauthoryear{Jorstad \etal}{2010}]{jorstad:10}Jorstad,
  S. G. \etal\ 2010, \apj, 715, 362.
%\bibitem[Just \etal(2007)]{2007ApJ...665.1004J} Just, D.~W., Brandt, 
%W.~N., Shemmer, O., et al.\ 2007, \apj, 665, 1004 
\bibitem[\protect\citeauthoryear{Kataoka \etal}{2000}]{kataoka:00} Kataoka,
  J., Takahashi, T., Makino, F., Madejski G. M., Tashiro M., Urry C. M., Kubo
  H., 2000, \apj, 528, 243
\bibitem[\protect\citeauthoryear{Landt \etal}{2008}]{landt:08} Landt, H., Padovani, P., Giommi, P.,
  Perri, M., \& Cheung, C.~C.\ 2008, \apj, 676, 87
\bibitem[\protect\citeauthoryear{Lazarova \etal}{2012}]{lazarova:12} Lazarova, M.~S., Canalizo, G.,
  Lacy, M., \& Sajina, A.\ 2012, \apj, 755, 29
\bibitem[\protect\citeauthoryear{Lewis \etal}{2003}]{lewis:03} Lewis, G.~F., Chapman, S.~C., \& Kuncic,
Z.\ 2003, \apjl, 596, L35
\bibitem[\protect\citeauthoryear{Marscher \etal}{2008}]{marscher:08}Marscher, A. P., \etal\ 2008, Nature, 452, 966
\bibitem[\protect\citeauthoryear{Marscher \etal}{2010}]{marscher:10}Marscher, A. P., \etal\ 2010, \apj, 710, L126
\bibitem[\protect\citeauthoryear{Marshall \etal}{2005}]{marshall:05} Marshall, H.~L., \etal\ 2005,
  \apjs, 156, 13
\bibitem[\protect\citeauthoryear{Miller \etal}{2006}]{miller:06} Miller, B.~P., Brandt, W.~N.,
Gallagher, S.~C., Laor, A., Wills, B.~J., Garmire, G.~P., \& Schneider, D.~P.\
2006, \apj, 652, 163
\bibitem[\protect\citeauthoryear{Miller \etal}{2009}]{miller:09} Miller, B.~P., Brandt, 
W.~N., Gibson, R.~R., Garmire, G.~P., \& Shemmer, O.\ 2009, \apj, 702, 911
\bibitem[\protect\citeauthoryear{Morrison \& McCammon}{1983}]{morrison:83} Morrison, R., \& McCammon,
D.\ 1983, \apj, 270, 119
\bibitem[\protect\citeauthoryear{Najita \etal}{2000}]{najita:00} Najita, J., Dey, A., \& Brotherton,
M.\ 2000, \aj, 120, 2859
\bibitem[\protect\citeauthoryear{Page \etal}{2005}]{page:05} Page, K.~L., Reeves, J.~N., O'Brien,
P.~T., \& Turner, M.~J.~L.\ 2005, \mnras, 364, 195
%\bibitem[\protect\citeauthoryear{Priddey \etal}{2007}]{priddey:07} Priddey, R.~S., Gallagher, S.~C.,
%Isaak, K.~G., Sharp, R.~G., McMahon, R.~G., \& Butner, H.~M.\ 2007, \mnras,
%374, 867
\bibitem[\protect\citeauthoryear{Punsly}{2006}]{punsly:06} Punsly, B.\ 2006,
  \apj, 647, 886
\bibitem[Reynolds \etal(2013)]{reynolds:13} Reynolds, C., Punsly, B., \&
  O'Dea, C.~P.\ 2013, \apjl, 773, L10
\bibitem[\protect\citeauthoryear{Richards \etal}{2006}]{richards:06} Richards, G.~T., \etal\ 2006,
\apjs, 166, 470
\bibitem[\protect\citeauthoryear{Sambruna \etal}{2007}]{sambruna:07} Sambruna, R.~M., Donato, D.,
  Tavecchio, F., Maraschi, L., Cheung, C.~C., \& Urry, C.~M.\ 2007, \apj, 670,
  74
\bibitem[\protect\citeauthoryear{Sambruna \etal}{2006}]{sambruna:06} Sambruna, R.~M., Gliozzi, M.,
  Donato, D., Maraschi, L., Tavecchio, F., Cheung, C.~C., Urry, C.~M., \&
  Wardle, J.~F.~C.\ 2006, \apj, 641, 717
%\bibitem[\protect\citeauthoryear{Schaefer \etal}{2006}]{schaefer:06} Schaefer, J.~J., Brotherton,
%M.~S., Shang, Z., Gregg, M.~D., Becker, R.~H., Laurent-Muehleisen, S.~A.,
%Lacy, M., \& White, R.~L.\ 2006, \aj, 132, 1464
\bibitem[\protect\citeauthoryear{Shankar \etal}{2008}]{shankar:08} Shankar, F., Dai, X., \& Sivakoff,
  G.~R.\ 2008, \apj, 687, 859
%\bibitem[\protect\citeauthoryear{Sprayberry \& Foltz}{1992}]{sprayberry:92} Sprayberry, D., \& Foltz,
%  C.~B.\ 1992, \apj, 390, 39
\bibitem[\protect\citeauthoryear{Steffen \etal}{2006}]{steffen:06} Steffen, A.~T., Strateva, I.,
Brandt, W.~N., Alexander, D.~M., Koekemoer, A.~M., Lehmer, B.~D., Schneider,
D.~P., \& Vignali, C.\ 2006, \aj, 131, 2826
\bibitem[\protect\citeauthoryear{Stocke \etal}{1992}]{stocke:92} Stocke, J.~T., Morris, S.~L., Weymann,
R.~J., \& Foltz, C.~B.\ 1992, \apj, 396, 487
\bibitem[\protect\citeauthoryear{Streblyanska \etal}{2010}]{streblyanska:10} Streblyanska, A., Barcons,
  X., Carrera, F.~J., \& Gil-Merino, R.\ 2010, \aap, 515, A2
\bibitem[\protect\citeauthoryear{Takahashi \etal}{1996}]{takahashi:96}
  Takahashi, T., \etal\ 1996, \apj, 470, L89
\bibitem[\protect\citeauthoryear{Tozzi \etal}{2006}]{tozzi:06} Tozzi, P., \etal\ 2006, \aap, 451, 457
\bibitem[\protect\citeauthoryear{Wang \etal}{2006}]{wang:06} Wang, T.-G., Zhou, H.-Y., Wang, J.-X., Lu,
Y.-J., \& Lu, Y.\ 2006, \apj, 645, 856
\bibitem[\protect\citeauthoryear{Wang \etal}{2008}]{wang:08} Wang, J., Jiang,
  P., Zhou, H., Wang T., Dong X, Wang H., 2008, \apjl, 676, L97
%\bibitem[\protect\citeauthoryear{Wills \etal}{1999}]{wills:99} Wills, B.~J., Brandt, W.~N., \& Laor,
%A.\ 1999, \apjl, 520, L91
\bibitem[\protect\citeauthoryear{York \etal}{2000}]{york:00} York,
  D.~G. \etal\ 2000, \aj, 120, 1579
\bibitem[\protect\citeauthoryear{Zhou \etal}{2006}]{zhou:06} Zhou, H., Wang, T., Wang, H., Wang, J.,
Yuan, W., \& Lu, Y.\ 2006, \apj, 639, 716
\end{thebibliography}
\end{document}